\documentclass[12pt]{emulateapj}
\usepackage{natbib}

\newcommand{\eg}{e.g.,}

\newcommand{\ie}{i.e.,}
\newcommand{\spitzer}{{\it Spitzer}}

\newcommand{\hmpc}{\hbox{$h^{-1}$ Mpc}}

\newcommand{\msun}{$M_\odot$}

\newcommand{\clname}{SPT-CL~J0546-5345}
\newcommand{\rfive}{r_{\mbox{\scriptsize 500}}}
\newcommand{\rtwo}{r_{\mbox{\scriptsize 200}}}

\newcommand{\Mfive}{M_{\mbox{\scriptsize 500}}}
\newcommand{\Mtwo}{M_{\mbox{\scriptsize 200}}}

\newcommand{\Mgfive}{M_{\mbox{\scriptsize g,500}}}
\newcommand{\Mfiveyx}{$M_{\mbox{\scriptsize 500,$Y_X$}}$}
\newcommand{\Ysz}{Y_{\mbox \scriptsize SZ}}

\newcommand{\Mtwoyx}{$M_{\mbox{\scriptsize 200,$Y_X$}}$}
\newcommand{\Mtwotx}{$M_{\mbox{\scriptsize 200,$T_X$}}$}
\newcommand{\MytwoSZE}{$M_{\mbox{\scriptsize 200,SZ}}$}
\newcommand{\MtwoYSZ}{$M_{\mbox{\scriptsize 200,$Y_{\mbox{SZ}}$}}$}

\newcommand{\Ntwo}{N_{\mbox{\scriptsize 200}}}
\newcommand{\Ntwobf}{N_{\mbox{\scriptsize \bf 200}}}
\newcommand{\Ngal}{N_{\mbox{\scriptsize gal}}}
\newcommand{\sigmadm}{\sigma_{\mbox{\scriptsize DM}}}
\newcommand{\sigmadmfifteen}{\sigma_{\mbox{\scriptsize DM,15}}}

\def\spose#1{\hbox to 0pt{#1\hss}}
\def\simlt{\mathrel{\spose{\lower 3pt\hbox{$\mathchar"218$}}
     \raise 2.0pt\hbox{$\mathchar"13C$}}}
\def\simgt{\mathrel{\spose{\lower 3pt\hbox{$\mathchar"218$}}
     \raise 2.0pt\hbox{$\mathchar"13E$}}}

\slugcomment{ApJ, in press}

 \shorttitle{\clname: A Massive $z>1$ Galaxy Cluster Selected Via the
   SZE}

\newcommand{\CfA}{1}
\newcommand{\KeckFellow}{2}
\newcommand{\Harvard}{3}
\newcommand{\Cardiff}{4}
\newcommand{\UChicago}{5}
\newcommand{\MIT}{6}
\newcommand{\Munich}{7}
\newcommand{\ExcellenceCluster}{8}
\newcommand{\KICPChicago}{9}
\newcommand{\EFIChicago}{10}
\newcommand{\PhysicsUChicago}{11}
\newcommand{\AAUChicago}{12}
\newcommand{\McGill}{13}
\newcommand{\Illinois}{14}
\newcommand{\PennState}{15}
\newcommand{\Berkeley}{16}
\newcommand{\UFlorida}{17}
\newcommand{\Colorado}{18}
\newcommand{\NASA}{19}
\newcommand{\Davis}{20}
\newcommand{\LBNL}{21}
\newcommand{\MPE}{22}
\newcommand{\Michigan}{23}
\newcommand{\CaseWestern}{24}
\newcommand{\Yale}{25}
\newcommand{\ClayFellow}{26}
\newcommand{\CTIOVisiting}{27}

\begin{document}


\title{\clname: A Massive $z>1$ Galaxy Cluster Selected Via the
  Sunyaev-Zel'dovich Effect with the South Pole Telescope}


\author{M.~Brodwin,\altaffilmark{\CfA,\KeckFellow,\CTIOVisiting}
J.~Ruel,\altaffilmark{\Harvard}
P.~A.~R.~Ade,\altaffilmark{\Cardiff}
K.~A.~Aird,\altaffilmark{\UChicago}
K.~Andersson,\altaffilmark{\MIT}
M.~L.~N.Ashby,\altaffilmark{\CfA} 
M.~Bautz,\altaffilmark{\MIT}
G.~Bazin,\altaffilmark{\Munich,\ExcellenceCluster}
B.~A.~Benson,\altaffilmark{\KICPChicago,\EFIChicago}
L.~E.~Bleem,\altaffilmark{\KICPChicago,\PhysicsUChicago}
J.~E.~Carlstrom,\altaffilmark{\KICPChicago,\EFIChicago,\PhysicsUChicago,\AAUChicago}
C.~L.~Chang,\altaffilmark{\KICPChicago,\EFIChicago}
T.~M.~Crawford,\altaffilmark{\KICPChicago,\AAUChicago}
A.~T.~Crites,\altaffilmark{\KICPChicago,\AAUChicago}
T.~de~Haan,\altaffilmark{\McGill}
S.~Desai,\altaffilmark{\Illinois}
M.~A.~Dobbs,\altaffilmark{\McGill}
J.~P.~Dudley,\altaffilmark{\McGill}
G.~G.~Fazio,\altaffilmark{\CfA}
R.~J.~Foley,\altaffilmark{\CfA,\ClayFellow}
W.~R.~Forman,\altaffilmark{\CfA}
G.~Garmire,\altaffilmark{\PennState}
E.~M.~George,\altaffilmark{\Berkeley}
M.~D.~Gladders,\altaffilmark{\KICPChicago,\AAUChicago}
A.~H.~Gonzalez,\altaffilmark{\UFlorida}
N.~W.~Halverson,\altaffilmark{\Colorado}
F.~W.~High,\altaffilmark{\Harvard}
G.~P.~Holder,\altaffilmark{\McGill}
W.~L.~Holzapfel,\altaffilmark{\Berkeley}
J.~D.~Hrubes,\altaffilmark{\UChicago}
C.~Jones,\altaffilmark{\CfA}
M.~Joy,\altaffilmark{\NASA}
R.~Keisler,\altaffilmark{\KICPChicago,\PhysicsUChicago}
L.~Knox,\altaffilmark{\Davis}
A.~T.~Lee,\altaffilmark{\Berkeley,\LBNL}
E.~M.~Leitch,\altaffilmark{\KICPChicago,\AAUChicago}
M.~Lueker,\altaffilmark{\Berkeley}
D.~P.~Marrone,\altaffilmark{\UChicago,\KICPChicago}
J.~J.~McMahon,\altaffilmark{\KICPChicago,\EFIChicago,\Michigan}
J.~Mehl,\altaffilmark{\KICPChicago,\AAUChicago}
S.~S.~Meyer,\altaffilmark{\KICPChicago,\EFIChicago,\PhysicsUChicago,\AAUChicago}
J.~J.~Mohr,\altaffilmark{\Munich,\ExcellenceCluster,\MPE,\CTIOVisiting}
T.~E.~Montroy,\altaffilmark{\CaseWestern}
S.~S.~Murray,\altaffilmark{\CfA}
S.~Padin,\altaffilmark{\KICPChicago,\AAUChicago}
T.~Plagge,\altaffilmark{\AAUChicago,\Berkeley}
C.~Pryke,\altaffilmark{\KICPChicago,\EFIChicago,\AAUChicago}
C.~L.~Reichardt,\altaffilmark{\Berkeley}
A.~Rest,\altaffilmark{\Harvard}
J.~E.~Ruhl,\altaffilmark{\CaseWestern}
K.~K.~Schaffer,\altaffilmark{\KICPChicago,\EFIChicago}
L.~Shaw,\altaffilmark{\McGill,\Yale}
E.~Shirokoff,\altaffilmark{\Berkeley}
J.~Song,\altaffilmark{\Illinois}
H.~G.~Spieler,\altaffilmark{\LBNL}
B.~Stalder,\altaffilmark{\CfA}
S.~A.~Stanford,\altaffilmark{\Davis}
Z.~Staniszewski,\altaffilmark{\CaseWestern}
A.~A.~Stark,\altaffilmark{\CfA}
C.~W.~Stubbs,\altaffilmark{\CfA,\Harvard}
K.~Vanderlinde,\altaffilmark{\McGill}
J.~D.~Vieira,\altaffilmark{\KICPChicago,\PhysicsUChicago}
A. Vikhlinin,\altaffilmark{\CfA}
R.~Williamson,\altaffilmark{\KICPChicago,\AAUChicago}
Y.~Yang,\altaffilmark{\Illinois}
O.~Zahn,\altaffilmark{\Berkeley}
and
A.~Zenteno\altaffilmark{\Munich,\ExcellenceCluster}
}


\altaffiltext{\CfA}{Harvard-Smithsonian Center for Astrophysics, 60 Garden Street, Cambridge, MA 02138}
\altaffiltext{\KeckFellow}{W. M. Keck Postdoctoral Fellow at the Harvard-Smithsonian Center for Astrophysics}
\altaffiltext{\Harvard}{Department of Physics, Harvard University, 17 Oxford Street, Cambridge, MA 02138}
\altaffiltext{\Cardiff}{Department of Physics and Astronomy, Cardiff University, CF24 3YB, UK}
\altaffiltext{\UChicago}{University of Chicago, 5640 South Ellis Avenue, Chicago, IL 60637}
\altaffiltext{\MIT}{Kavli Institute for Astrophysics and Space Research, MIT, Cambridge, MA 02139}
\altaffiltext{\Munich}{Department of Physics, Ludwig-Maximilians-Universit\"{a}t, Scheinerstr.\ 1, 81679 M\"{u}nchen, Germany}
\altaffiltext{\ExcellenceCluster}{Excellence Cluster Universe, Boltzmannstr.\ 2, 85748 Garching, Germany}
\altaffiltext{\KICPChicago}{Kavli Institute for Cosmological Physics, University of Chicago, 5640 South Ellis Avenue, Chicago, IL 60637}
\altaffiltext{\EFIChicago}{Enrico Fermi Institute, University of Chicago, 5640 South Ellis Avenue, Chicago, IL 60637}
\altaffiltext{\PhysicsUChicago}{Department of Physics, University of Chicago, 5640 South Ellis Avenue, Chicago, IL 60637}
\altaffiltext{\AAUChicago}{Department of Astronomy and Astrophysics, University of Chicago, 5640 South Ellis Avenue, Chicago, IL 60637}
\altaffiltext{\McGill}{Department of Physics, McGill University, 3600 Rue University, Montreal, Quebec H3A 2T8, Canada}
\altaffiltext{\Illinois}{Department of Astronomy, University of Illinois, 1002 West Green Street, Urbana, IL 61801}
\altaffiltext{\PennState}{Department of Astronomy and Astrophysics, Pennsylvania State University, 525 Davey Lab, University Park, PA 16802, USA}
\altaffiltext{\Berkeley}{Department of Physics, University of California, Berkeley, CA 94720}
\altaffiltext{\UFlorida}{Department of Astronomy, University of Florida, Gainesville, FL 32611}
\altaffiltext{\Colorado}{Department of Astrophysical and Planetary Sciences and Department of Physics, University of Colorado, Boulder, CO 80309}
\altaffiltext{\NASA}{Department of Space Science, VP62, NASA Marshall Space Flight Center, Huntsville, AL 35812}
\altaffiltext{\Davis}{Department of Physics, University of California, One Shields Avenue, Davis, CA 95616}
\altaffiltext{\LBNL}{Physics Division, Lawrence Berkeley National Laboratory, Berkeley, CA 94720}
\altaffiltext{\MPE}{Max-Planck-Institut f\"{u}r extraterrestrische Physik, Giessenbachstr.\ 85748 Garching, Germany}
\altaffiltext{\Michigan}{Department of Physics, University of Michigan, 450 Church Street, Ann Arbor, MI, 48109}
\altaffiltext{\CaseWestern}{Physics Department and CERCA, Case Western Reserve University, 10900 Euclid Ave., Cleveland, OH 44106}
\altaffiltext{\Yale}{Department of Physics, Yale University, P.O. Box 208210, New Haven, CT 06520-8120}
\altaffiltext{\ClayFellow}{Clay Fellow}
\altaffiltext{\CTIOVisiting}{Visiting astronomer, Cerro Tololo Inter-American
  Observatory, National Optical Astronomy Observatory, under contract
  with the National Science Foundation.}


\begin{abstract}

  We report the spectroscopic confirmation of \clname\ at $\left< z
  \right> = 1.067$.  To date this is the most distant cluster to be
  spectroscopically confirmed from the 2008 South Pole Telescope (SPT)
  catalog, and indeed the first $z>1$ cluster discovered by the
  Sunyaev-Zel'dovich Effect (SZE).  We identify 21 secure
  spectroscopic members within $0.9$ Mpc of the SPT cluster position,
  18 of which are quiescent, early-type galaxies.  From these
  quiescent galaxies we obtain a velocity dispersion of $1179\,
  ^{+232}_{-167}\,$ km/s, ranking \clname\ as the most dynamically
  massive cluster yet discovered at $z>1$.  Assuming that \clname\ is
  virialized, this implies a dynamical mass of $\Mtwo =
  1.0\,^{+0.6}_{-0.4} \times 10^{15}$ \msun, in agreement with the
  X-ray and SZE mass measurements.  Combining masses from several
  independent measures leads to a best-estimate mass of $\Mtwo = (7.95
  \pm 0.92) \times 10^{14}$ \msun.  The spectroscopic confirmation of
  \clname, discovered in the wide-angle, mass-selected SPT cluster
  survey, marks the onset of the high redshift SZE-selected galaxy
  cluster era.

\end{abstract}




\section{Introduction}

Formed from the extreme peaks of the primordial density field, galaxy
clusters are both a sensitive probe of cosmology and an excellent
laboratory for galaxy evolution studies. Although massive galaxy
clusters are easy to identify in the local Universe, finding such
objects at large lookback times has been extremely challenging. A
patchwork of optical, infrared and X-ray techniques has led to a
heterogeneous sample of distant galaxy clusters
\citep[\eg][]{stanford97,rosati04, gladders05, mullis05,
  brodwin06_ISS, elston06, stanford06, eisenhardt08, muzzin09,
  papovich10}.  While each technique offers unique advantages, namely,
observational efficiency (optical), redshift reach and mass
sensitivity (IR), and a direct mass-observable (X-ray), none produce
purely mass-selected, nearly redshift independent cluster samples.

A long-awaited advance is the arrival of Sunyaev-Zel'dovich Effect
(SZE) cluster surveys.  The SZE is a small distortion in the Cosmic
Microwave Background (CMB) spectrum caused by the inverse Compton
scattering of CMB photons with the hot intracluster medium
\citep[ICM;][]{sunyaev72}.  The surface brightness of the SZE does not
depend on a cluster's distance, which makes it a nearly
redshift-independent method for finding massive galaxy clusters
\citep[\eg][]{carlstrom02}.  A new generation of dedicated
millimeter-wave (mm-wave) experiments to search for galaxy clusters
are currently surveying large areas of sky, including the Atacama
Cosmology Telescope \citep{fowler07} and the South Pole Telescope
\citep{carlstrom09}.

Taking advantage of the excellent conditions for mm-wave observations
at the South Pole, the 10-m South Pole Telescope
\citep[SPT;][]{carlstrom09} is midway through a $\gtrsim 2000$ deg$^2$
survey sensitive to galaxy clusters above $\approx 5 \times 10^{14}$
\msun\ at all redshifts.  The first three clusters discovered by the
SZE were reported by \citet[][hereafter S09]{zak09}.  The complete 5
$\sigma$ cluster catalog from the 2008 SPT season, a study of the
optical properties of these clusters, and a description of the survey
strategy and goals are given in \citet[][hereafter V10]{vanderlinde10}
and \citet[][hereafter H10]{high10}.

Here we report the spectroscopic confirmation, velocity dispersion and
dynamical mass of the first $z>1$ SZE-selected cluster, \clname, at
$\left< z \right> = 1.067$.  In converting from $\rfive$ to $\rtwo$
and $\Mfive$ to $\Mtwo$ throughout, we assume an NFW density profile
and the mass-concentration relation of \citet{duffy08}.  Unless
otherwise indicated, we use Vega magnitudes and adopt a WMAP5
cosmology \citep{dunkley09}, with $\Omega_M = 0.264$, $\Omega_\Lambda
= 0.736$, $\sigma_{\mbox{\scriptsize 8}} = 0.80$ and $H_0 = 71$ km/s/Mpc.

\section{Observations}
\label{Sec: data}

\subsection{SPT-SZE}

In 2008, the SPT survey observed a 196 deg$^2$ area (178 deg$^2$ after
point-source and non-uniform coverage masking) for $\sim 1500$ hr
(V10), reaching a 5 $\sigma$ point source sensitivity of $\sim 7$ mJy
at 150 GHz (2.0 mm).  Details of the data processing, map-making,
cluster extraction and significance estimates are given in S09 and
V10. \clname, first reported in S09, has an SZE decrement of S/N =
7.69 at 150 GHz.  It is among the most significant of the 2008 cluster
candidates with a $z \ga 1$ optical red-sequence (H10). SZE
significance contours at 150 GHz are shown in Figure \ref{Fig: image}
({\it left}).

\subsection{\spitzer/IRAC and Optical Imaging}

Mid-infrared \spitzer/IRAC imaging was obtained in 2009 September as
part of a larger program to follow up clusters identified in the SPT
survey (PID 60099; PI Brodwin). The on-target observations consisted
of 8$\times100$\,s and 6$\times30$\,s dithered exposures at 3.6 and
$4.5\,\mu$m, respectively. The deep 3.6\,$\mu$m observations are
sensitive to passively-evolving cluster galaxies down to 0.1 L$^*$ at
$z = 1.5$.  The data were reduced following the method of
\citet{ashby09}.  Briefly, we correct for column pulldown, mosaic the
individual exposures, resample to 0\farcs86 pixels (half the solid
angle of the native IRAC pixels), and reject cosmic rays.

Optical data were provided by the Blanco Cosmology Survey (BCS), an
NOAO survey program with deep $griz$ imaging over 100\,deg$^2$ in a
subset of the SPT 05hr and 23hr fields \citep{bcs}.  The Stellar Locus
Regression method of \citet{high09} was used for photometric
calibration, extinction correction and Galactic de-reddening.  The
independent calibration from the DES data management system
\citep{mohr08} gave consistent results.  The BCS data typically reach
5 $\sigma$ AB depths of 24.75, 24.65, 24.35, and 23.5 in $griz$.  See
S09 and H10 for a more detailed description of the optical data.

\begin{figure*}[hbpt]
\plotone{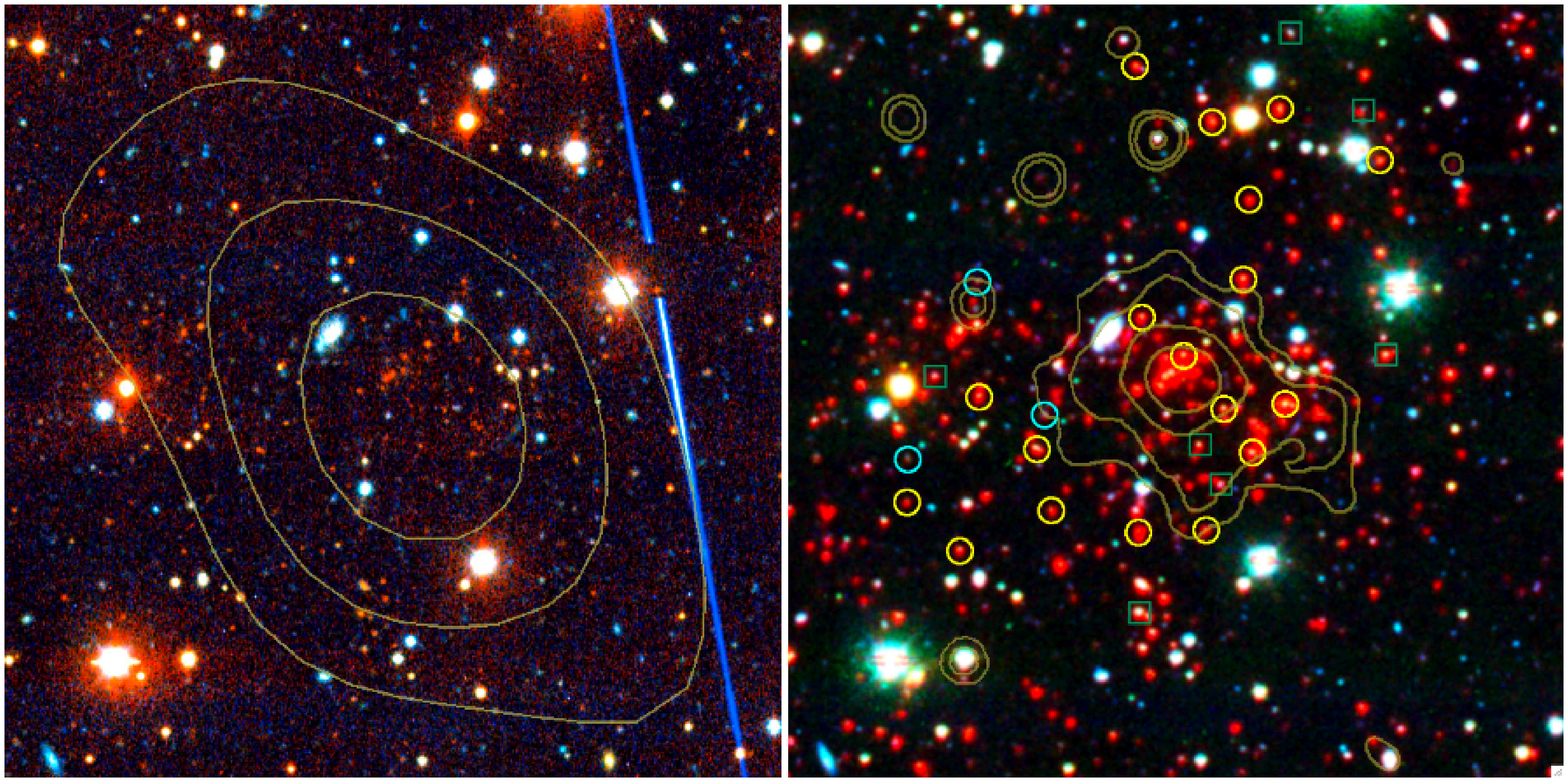}
\caption{{\it Left:} Optical $4\arcmin \times 4\arcmin$ color image
  ($grz$) of \clname, with SZE significance contours overlaid (S/N =
  2, 4, and 6).  {\it Right:} False color optical ($ri$)$\,+\,$IRAC
  ($3.6\,\mu$m) image of \clname, with {\it Chandra} X-ray contours
  overlaid (0.25, 0.4, 0.85 and 1.6 counts per $2\arcsec \times
  2\arcsec$ pixel per 55.6 ks in the 0.5-2 keV band).  North is up,
  east is to the left.  Due to its high angular resolution, {\it
    Chandra} is able to resolve substructure to the SW, which may be
  evidence of a possible merger.  These images highlight the
  importance of IRAC imaging in studying the galaxies in high
  redshift, optically faint clusters.  Spectroscopic early-type
  (late-type) members are indicated with yellow (cyan) circles.  Green
  squares show the spectroscopic non-members.}
\label{Fig: image}
\end{figure*}

Along with the $r$ and $i$ BCS photometry, the IRAC 3.6\,$\mu$m data
were used to select promising cluster member candidates for
spectroscopy as described below. A false-color optical
($ri$)$\,+\,$IRAC ($3.6\,\mu$m) image is shown in Figure \ref{Fig:
  image} ({\it right}). The optical-only ({\it grz)} image (Figure
\ref{Fig: image}, {\it left}) indicates how optically (rest-frame UV)
faint the cluster members are, and highlights the importance of
\spitzer/IRAC imaging for high-redshift cluster detection and study.

\subsection{\it Chandra}

\clname\ was observed with {\it Chandra}/ACIS-I on four separate
occasions, for a total exposure time of 55.6 ks (PID 09800046; PI
Garmire, PID 11800471; PI Mohr). This yielded 1304 source counts in
the $0.5-7.0$ keV energy range within $0.5 \, \rfive$. Background was
subtracted using blank sky data normalized to the flux in the $9.5-12$
keV energy band.  The cluster luminosity is $L_X (0.5-2\, \mbox{keV})
= (6.4 \pm 0.4) \times 10^{44}$ erg s$^{-1}$, and the temperature,
estimated in the $(0.15-1)\, \rfive$ annulus, is $T_X = 7.5\,
^{+1.7}_{-1.1}\,$ keV.  The gas mass was estimated to be $\Mgfive =
7.3 \, ^{+0.4}_{-0.3}\, \times 10^{13}$ \msun\ from the X-ray surface
brightness in the $0.5-2.0$ keV band following the analysis method
described in \citep{vikhlinin06}, where the gas density is assumed to
follow a modified $\beta$-model \citep{cavaliere78} that is allowed to
steepen at large radii and to have a power-law cusp at the cluster
center.  The {\it Chandra} X-ray contours are overlaid in Figure
\ref{Fig: image} ({\it right}).  The reader is referred to
\citet[][hereafter A10]{andersson10} for a full description of the
X-ray analysis of \clname, as well as a study of the X-ray properties
of 15 of the SZE-selected clusters from the V10 sample.

\subsection{Optical Spectroscopy}

Multislit spectroscopic observations were acquired on the 6.5-meter
Magellan Baade telescope on UT 2010 February 11. We used the Gladders
Image-Slicing Multi-Slit Option
(GISMO\footnote{http://www.lco.cl/telescopes-information/magellan/\\instruments/imacs/gismo},
Gladders et al.~in prep) module on the Inamori Magellan Areal Camera
and Spectrograph \citep[IMACS;][]{dressler06}. GISMO optically remaps
the central region of the IMACS field-of-view (roughly $3.5\arcmin
\times 3.2\arcmin$) to sixteen evenly-spaced regions of the focal
plane, allowing for a large density of slitlets in the cluster core
while minimizing trace overlaps on the CCD. We used the f/2 camera for
its greater red sensitivity, the $300$ l/mm ``red'' grism and the
WB6300-9500 filter. The seeing was excellent ($\la 0.5\arcsec$)
throughout the six 30 min exposures.

In designing the multislit mask, galaxies were assigned a weight
proportional to their $i$-band brightness and their proximity to the
cluster center, and inversely proportional to their distance in color
space, in both $r-i$ and $i-[3.6]$, from the predicted color from a
\citet[][hereafter BC03]{bc03} passively evolving 100 Myr burst model
with formation redshift $z_f = 3$.  

The COSMOS reduction
package\footnote{http://obs.carnegiescience.edu/Code/cosmos/} was used
for standard CCD processing, resulting in wavelength-calibrated 2D
spectra.  The 1D spectra were then extracted from the sum of the
reduced data. We flux calibrated the data and removed telluric
absorption using the continuum of a spectrophotometric standard
\citep{foley03}.

\section{Results}

\subsection{Redshifts}

Redshifts for galaxy spectra exhibiting significant absorption
features were found through cross correlation using the RVSAO package
in IRAF \citep{kurtz98}. The validity of the cross-correlation
redshift was checked by visual inspection and judged by the presence
of visible absorption lines; most of these displayed clear Ca HK
lines.  For three members the redshift was determined from a strong,
unresolved [OII] emission line. Redshift uncertainties were estimated
as twice those given by RVSAO \citep{quintana00}.  An independent
extraction and visual redshift determination 
yielded identical redshifts for all cluster members within the errors.

\begin{figure}[hbpt]
\epsscale{1.1}
\plotone{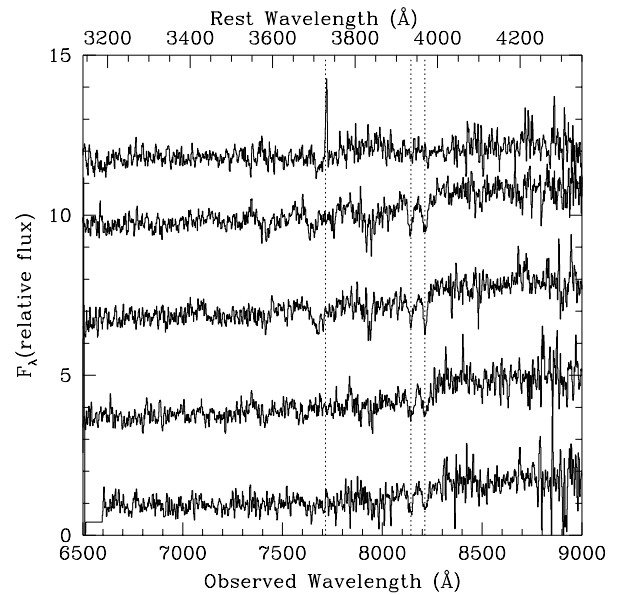}
\caption{Representative spectra of cluster member galaxies for
  \clname.  Vertical dotted lines indicate [OII] and Ca HK features.
  Most (18/21) confirmed members are passive, early-type galaxies.}
\label{Fig: spectra}
\end{figure}

From these Magellan observations we obtained 28 secure redshifts, of
which 21 are spectroscopic cluster members within $r < 0.9$ Mpc of the
SPT cluster position.  This high success rate validates our
spectroscopic selection algorithm.  Representative spectra are shown
in Figure \ref{Fig: spectra}.  The majority of the members (18) are
quiescent, with redshifts determined from Ca HK absorption, as
expected given their red-sequence selection, central location and the
extreme mass of the cluster.

\begin{deluxetable*}{lccllc}
  \tabletypesize{\normalsize} \tablecaption{Spectroscopic members of
    \clname\label{Tab: table}} \tablewidth{0pt} \tablehead{ \colhead{}
    & \colhead{R.A.} & \colhead{Dec.} & \colhead{} & \colhead{} &
    \colhead{Principal} \\
    \colhead{ID} & \colhead{(J2000)} & \colhead{(J2000)} &
    \colhead{$z$} & \colhead{$\delta z$\tablenotemark{a}} &
    \colhead{Spectral Feature}} \startdata
J054637.2-534525  &  05:46:37.23  & -53:45:25.6  & 1.0647  &  0.0002 & Ca HK/D4000  \\ 
J054636.2-534413  &  05:46:36.22  & -53:44:13.4  & 1.0656  &  0.0004 & Ca HK/D4000  \\ 
J054642.3-534554  &  05:46:42.36  & -53:45:54.5  & 1.0548  &  0.0003 & Ca HK/D4000  \\ 
J054633.6-534540  &  05:46:33.67  & -53:45:40.5  & 1.0775  &  0.0005 & Ca HK/D4000  \\ 
J054635.1-534502  &  05:46:35.14  & -53:45:02.0  & 1.0676  &  0.0003 & Ca HK/D4000  \\ 
J054638.9-534356  &  05:46:38.90  & -53:43:56.2  & 1.0743  &  0.0003 & Ca HK/D4000  \\ 
J054634.8-534555  &  05:46:34.81  & -53:45:55.6  & 1.0567  &  0.0002 & Ca HK/D4000  \\ 
J054635.8-534542  &  05:46:35.82  & -53:45:42.0  & 1.0710  &  0.0003 & Ca HK/D4000  \\ 
J054638.6-534513  &  05:46:38.68  & -53:45:13.7  & 1.0692  &  0.0003 & Ca HK/D4000  \\ 
J054644.3-534538  &  05:46:44.36  & -53:45:38.4  & 1.0619  &  0.0002 & Ca HK/D4000  \\ 
J054645.0-534625  &  05:46:45.04  & -53:46:25.9  & 1.0662  &  0.0003 & Ca HK/D4000  \\ 
J054641.8-534613  &  05:46:41.84  & -53:46:13.5  & 1.0712  &  0.0004 & Ca HK/D4000  \\ 
J054630.3-534425  &  05:46:30.37  & -53:44:25.3  & 1.0502  &  0.0005 & Ca HK/D4000  \\ 
J054638.8-534620  &  05:46:38.80  & -53:46:20.2  & 1.0681  &  0.0004 & Ca HK/D4000  \\ 
J054633.8-534409  &  05:46:33.86  & -53:44:09.3  & 1.0586  &  0.0005 & Ca HK/D4000  \\ 
J054646.8-534610  &  05:46:46.89  & -53:46:10.9  & 1.0625  &  0.0005 & Ca HK/D4000  \\ 
J054634.9-534437  &  05:46:34.92  & -53:44:37.5  & 1.0705  &  0.0005 & Ca HK/D4000  \\ 
J054636.4-534619  &  05:46:36.43  & -53:46:19.5  & 1.0805  &  0.0007 & Ca HK/D4000  \\ 
J054644.4-534502  &  05:46:44.42  & -53:45:02.7  & 1.0661  &  0.0025 & [OII]  \\ 
J054646.8-534557  &  05:46:46.87  & -53:45:57.6  & 1.0686  &  0.0002 & [OII]  \\ 
J054642.0-534543  &  05:46:42.08  & -53:45:43.9  & 1.0524  &  0.0002 & [OII]  \\ 
\enddata
\tablenotetext{a}{Redshift errors are twice those given by RVSAO.}
\end{deluxetable*}

\begin{figure*}[hbpt]
\epsscale{1.1}
\plotone{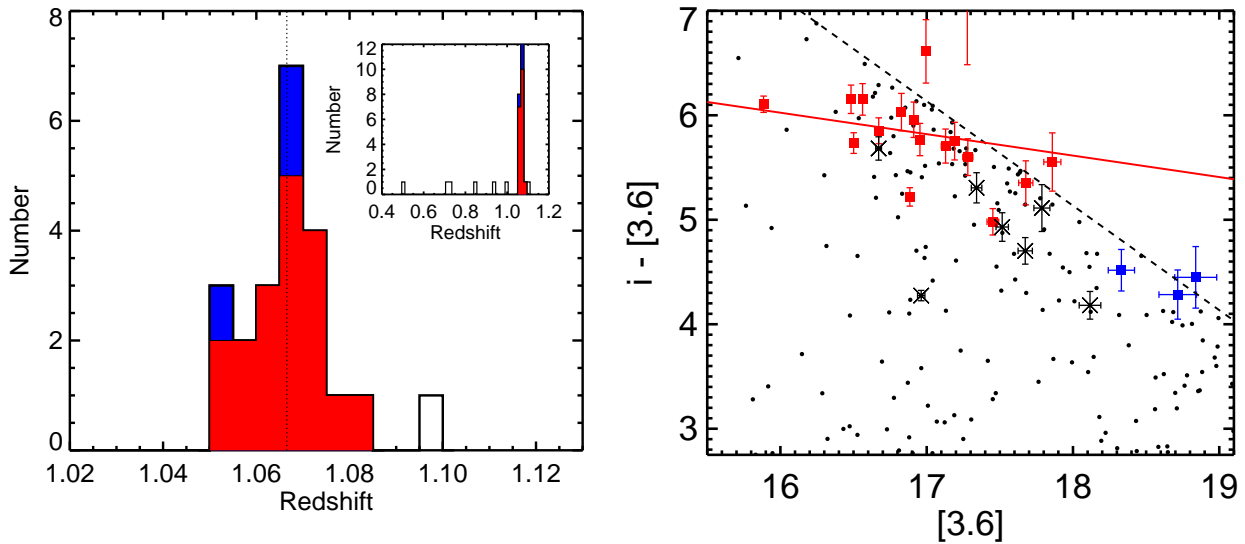}
\caption{{\it Left}: Histogram of spectroscopic redshifts obtained for
  \clname, showing a very broad peak consisting of 21 members (shaded
  regions) centered at $\left<z\right> = 1.0665$ (dotted line).  The
  passive galaxies used in the dynamical analysis are shaded red.
  Galaxies for which redshifts were determined from [OII] emission are
  shaded blue. Spectroscopic non-members are unshaded.  The inset
  shows all the secure redshifts obtained toward this cluster. {\it
    Right}: Color-magnitude diagram in $i-[3.6]$ within a 2.5\arcmin\
  radius around \clname.  The red (blue) squares are the
  spectroscopically confirmed passive ([OII]) members.  The passive
  members form a rich red-sequence.  Spectroscopic non-members are
  marked with X symbols.  The 5 $\sigma$ color-mag limit is indicated
  by the dashed line.  One member is omitted due to low S/N
  photometry.  The red line is the rest-frame Coma CMR, in the
  approximately $(U-H)$ color probed by our filters, normalized to
  match a BC03 passively evolving $L^*$ galaxy.}
\label{Fig: histogram}
\end{figure*}

\subsection{Velocity Dispersion}
\label{Sec: dispersion}

An iterative 3 $\sigma$ clipping algorithm was used to identify
cluster members.  We list the 21 secure cluster members in Table
\ref{Tab: table}, and plot the redshift histogram in Figure \ref{Fig:
  histogram} ({\it left}). In this Figure the 18 passive members, for
which the redshift was obtained from Ca HK absorption lines, are
shaded red. Redshifts of the remaining three members, shaded blue,
were secured from strong [OII] emission. Spectroscopic non-members are
unshaded. From the spectroscopic redshift histogram, \clname\ appears
to be a single massive halo, with no evidence of substructure or
merger activity.  We note, however, that the X-ray image (Figure 1)
shows a substructure extending to the SW indicating that a minor
merger may be taking place (A10).

We use the robust biweight estimator of \citet{beers90} to estimate
the mean velocity and velocity dispersion, applying the relativistic
correction and the usual correction for velocity errors
\citep{danese80}. For the full membership, we find a mean redshift and
velocity dispersion of $z = 1.0661 \,^{+0.0018}_{-0.0022}$ and $\sigma
= 1181\, ^{+215}_{-186}\,$ km/s, respectively.  The errors in both
quantities, obtained from bootstrap resampling, represent the 68\%
confidence interval.

Since late-type members are often infalling, they tend to yield
broader dispersions than early-types \citep[\eg][]{girardi96,fadda96,
  mohr96,koranyi00}.  We therefore explore limiting our analysis to
the 18 early-type members, as these are expected to better reflect the
cluster potential. This approach yields nearly identical results, with
a mean redshift of $z = 1.0665\, ^{+0.0019}_{-0.0021}$ and dispersion
of $\sigma = 1179\, ^{+232}_{-167}\, $ km/s.  Although the biweight
estimator is optimal, we also compute dispersions for these 18 members
using both the gapper method \citep{beers90} and the simple standard
deviation.  These yield $\sigma_{\mbox{\tiny gap}} = 1170\,
^{+240}_{-128}\,$ km/s and $\sigma_{\mbox{\tiny SD}} = 1138\,
^{+205}_{-132}\,$ km/s, respectively, both in excellent agreement with
the biweight dispersion.  There is no evidence of merger activity in
\clname\ from the line-of-sight velocities, in either the appearance
of Figure \ref{Fig: histogram} or in the dispersion measures, although
our redshift sampling is too sparse to rule it out.

\subsection{Color-Magnitude Relation}

We plot the $i-[3.6]$ color-magnitude diagram for galaxies within a
2.5\arcmin\ (1.22 Mpc) radius of \clname\ in Figure \ref{Fig:
  histogram} ({\it right}). The spectroscopically confirmed passive
galaxies define a rich red sequence, indicative of a cluster that is
already well-evolved at $z=1.067$. The dashed line indicates the
5$\sigma$ color-mag limit; one confirmed member is not detected
robustly enough in the optical for inclusion in this plot.  The solid
red line indicates the slope of the rest-frame color-magnitude
relation (CMR) of Coma \citep{eisenhardt07}, normalized to L$^*$ at
3.6\,$\mu$m in the passively evolving $z_f = 3$ BC03 model described
above.

\subsection{Dynamical Mass}
\label{Sec: dynamics}

There are several approaches in the literature for calculating the
dynamical mass.  Perhaps the most widely adopted is that of
\citet{carlberg97}, who use the simple definition $\Mtwo = 4/3\,
\pi\,\rtwo^3 \, 200\,\rho_c$, where $\rtwo = \sqrt{3}\, \sigma / 10\,
H(z)$ and $\rho_c$ is the critical density.  Although this approach
provides a convenient basis for comparison with other studies, it
relies on the simplifying assumption that clusters are singular
isothermal spheres.  We therefore prefer to use the simulation-based
$\sigma - \Mtwo$ relation of \citet{evrard08},
\begin{equation}
\Mtwo = \frac{10^{15}}{h(z)} \left( \frac{\sigmadm}{\sigmadmfifteen} \right) ^{1/\alpha},
\label{Eq: evrard}
\end{equation}
where $\sigmadmfifteen = 1082.9$ km/s, $\alpha = 0.3361$, and
$\sigmadm$, the dark matter velocity dispersion, is related to the
observed galaxy velocity dispersion by the velocity bias parameter,
$b_v \equiv \sigma/\sigmadm$.  Simulations generally suggest that $0.9
< b_v < 1.1$, and the most recent numerical results \citep[][and
references therein]{evrard08} indicate that galaxies are essentially
unbiased tracers of the dark matter potential (\ie\ $b_v = 1$).
Adopting this value, we obtain $\Mtwo = 1.04\,^{+0.61}_{-0.44} \times
10^{15}$ \msun.  Dynamical masses corresponding to the various
velocity dispersion measures reported in \textsection{\ref{Sec:
    dispersion}} are tabulated in Table \ref{Tab: masses}.  The
\citet{carlberg97} relation yields masses a factor of $\approx 1.5$
larger.

Detailed studies of the velocity dispersion profile
\citep[\eg][]{girardi93,fadda96} indicate that dispersions measured
within radii less than $\rtwo$ may be biased high.  In A10, $\rtwo$
for \clname\ is estimated to be $\approx 1.57$ Mpc, and we are
therefore probing an aperture of $\approx 0.57 \rtwo$.  Numerical
\citep{biviano06} and observational
\citep{rines06,katgert96,katgert98} studies suggest that for our
early-type galaxy sample and the clustercentric radius probed, the
bias is expected to be small.  For example, \citet{biviano06} report
the expected bias in the dispersion as a function of the fractional
virial radius probed; for our dispersion measurement the expected
upward bias is $\la 2.5\%$.  This implies a corresponding bias in the
dynamical mass of $\la 7\%$, which would lower the dynamical mass to
$\sim 9.6 \times 10^{14}$ \msun.  While this may well be a more
accurate value, the bias correction is both uncertain and small.  We
therefore defer a more extensive analysis to a future paper, when a
larger sample of spectroscopic members out to $\rtwo$ is available.

Finally, we note that no other $z>1$ cluster with $\ge 10$
spectroscopically confirmed members has $\sigma>1000$ km/s. With a
dispersion of $\sigma=1179$ km/s, \clname\ is unambiguously the most
dynamically massive cluster yet identified at $z>1$.

\section{Discussion}
\label{Sec: discussion}

\subsection{$\sigma - T_X$ Relation}
\label{Sec: sigma-tx}

\begin{figure}[hbpt]
\epsscale{1.2}
  \plotone{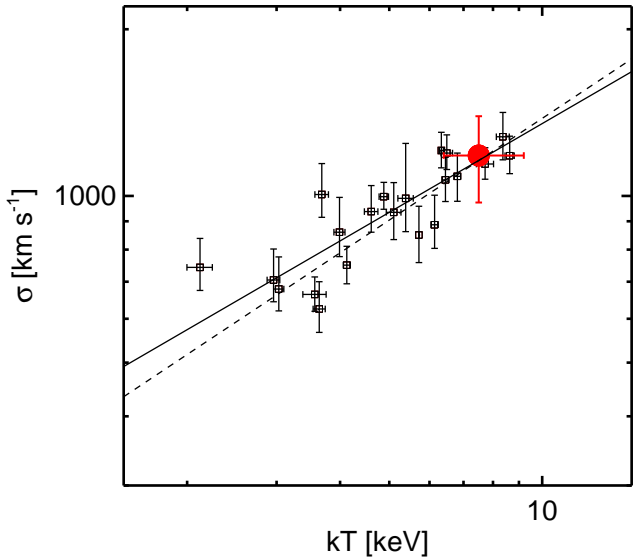}
  \caption{Empirical $\sigma - T_X$ correlation for the X-ray clusters
    from \citet{vikhlinin09}, with velocity dispersions taken from
    \citet{girardi96}. The temperatures are measured exactly as for
    \clname.  The fits, described in the text, are typical of those
    found by other authors \citep[\eg][]{lubin93,horner99}.  \clname\
    (red circle) falls right on this relation.}
  \label{Fig: sigma-Tx}
\end{figure}

At $T_X = 7.5 ^{+1.7}_{-1.1}$ keV, \clname\ is among the hottest
clusters yet observed in the X-ray at $z>1$ \citep{rosati09}.  Figure
\ref{Fig: sigma-Tx} shows the well-known correlation between X-ray
temperature and galaxy velocity dispersion \citep{lubin93, bird95,
  girardi96, horner99, xue00, ortiz-gil04}.  The clusters shown are
the subset of the well-studied, low redshift X-ray cluster sample of
\citet{vikhlinin09} for which accurate velocity dispersion
measurements, from at least 40 galaxies, are available from
\citet{girardi96}.  The temperatures are measured in an identical
manner to that of \clname, which is plotted as the large red
circle. The dashed line is the best-fit relation \citet{girardi96}
obtained using previous temperatures from the literature, and the
solid line is our own fit using the \citet{vikhlinin09}
temperatures. Both fits exclude \clname, and are consistent with
typical measurements \citep[\eg][]{lubin93,horner99}.  Although the
comparison clusters are all at low redshift ($z<0.1$), there is no
expectation of, nor evidence for, evolution in this relation to
$z\sim1$ \citep[\eg][]{wu98,tran99}. The measured dispersion and X-ray
temperature for \clname\ fall right on the $\sigma - T_X$ relation,
offering strong corroboration of the very large dynamical mass
presented here.

\subsection{ X-Ray, SZE and Optical Mass Measures}
\label{Sec: xray}

We now compare the dynamical mass with several other mass measures
available for \clname, all reported in terms of $\Mtwo (\rho_c)$.  A
summary of these these mass measures is given in Table \ref{Tab:
  masses}.  

In A10 the cluster mass within $\rfive$ is measured using the
$Y_X$--$\Mfive$ relation \citep{vikhlinin09} to be \Mfiveyx $\, =(5.33
\pm 0.62) \times 10^{14}$ \msun.  The systematic uncertainty of this
scaling relation calibration was estimated by \citet{vikhlinin09} to
be $\pm$9\% by comparing X-ray mass estimates to weak lensing mass
measurements for a representative sample of clusters.  We include this
uncertainty in quadrature when converting to \Mtwoyx, and we estimate
that \Mtwoyx $\, = (8.23 \pm 1.21) \times 10^{14}$ \msun.  This X-ray
mass is consistent with the measured dynamical mass within the
errors. Although the X-ray image shows evidence of a possible merger,
the $Y_X$ estimator is expected to be robust to such events
\citep{vikhlinin09}.  Indeed, the majority of high redshift clusters
have asymmetrical X-ray morphologies (\eg\ \citealt{vikhlinin09};
A10), so the inferred X-ray mass of \clname\ should be of comparable
accuracy to other clusters at similar redshifts.  We also estimate the
mass from the temperature, via the $T_X$--$\Mfive$ relation of
\citet{vikhlinin09}, to be \Mtwotx = $(8.11 \pm 1.89) \times 10^{14}$
\msun.

The SZE measurement of the integrated Comptonization, $\Ysz$, is
expected to be the most accurate SZ mass proxy \citep{motl05,nagai06}.
In A10, they measure a spherically deprojected Comptonization of $\Ysz
= (3.5 \pm 0.6) \times 10^{14}$ \msun keV, where the uncertainty is
estimated from the quadrature sum of statistical errors and the
uncertainty in the assumed gas profile.  From this measurement, we use
the $\Ysz$--$\Mfive$ relation reported in A10 to estimate a mass,
where we marginalize over the allowed range of normalization and slope
parameters from their fit.  We include an additional $\pm$9\%
uncertainty to account for the systematic uncertainty of the
$Y_X$--based masses used to calibrate their relation\footnote{The SZ
  and X-ray mass estimates aren't completely independent.  We expect
  that their largest correlation will be from the mass calibration of
  the $Y_X$--$\Mfive$ and $\Ysz$--$\Mfive$ relations, which are based
  on the same X-ray measurements from \citet{vikhlinin09}.  We add the
  9\% uncertainty to the $\Ysz$ mass estimate to account for this, and
  assume that any other correlated uncertainty is negligible.}
Using this method, we infer a mass of \MtwoYSZ $ = (7.19 \pm 1.51)
\times 10^{14}$ \msun, in agreement with the X-ray and dynamical mass
estimates.

Another SZ mass estimate can be inferred from the SZE
significance-mass scaling relation established via WMAP7-constrained
simulations in V10.  A10 compared this SZE significance-based mass
proxy with X-ray inferred masses for their sample of 15 SZE-selected
clusters, finding that the significance-based SZE masses may be biased
low, with an average ratio of the SZE to X-ray masses of $0.89 \pm
0.06$.  The result for \clname, \MytwoSZE $\, = (5.03 \pm 1.13 \pm
0.77) \times 10^{14}$ \msun, where the uncertainties are statistical
followed by systematic, is indeed $\sim 1.6 \sigma$ below the 68\%
confidence interval of the more precise X-ray mass, although it is in
agreement with the dynamical mass within the errors.  Accounting for
the bias quantified in A10 brings the significance-based mass estimate
into agreement with the X-ray mass.

H10 present a scaling relation between SZE-derived mass and two
optical richness measures, $\Ngal$ and $\Ntwo$.  These richness
measures count the number of galaxies on the red-sequence at the
cluster redshift within clustercentric radii of 1 \hmpc\ and $\rtwo$,
respectively.  The derived masses, $\Mtwo(\Ngal) = (9.2 \pm 4.9 \pm
2.7) \times \,10^{14}$ \msun\ and $\Mtwo(\Ntwo) = (8.5 \pm 5.7 \pm
2.5) \times \,10^{14}$ \msun, are consistent with the independent
richness-based mass estimate of \citet{menanteau09} for this cluster.
Although noisy mass proxies, these richness measures offer independent
evidence of the high mass of \clname.

\begin{deluxetable*}{llcclc}
\tablecaption{Comparison of Mass Measurements for \clname\label{Tab: masses}} \tablewidth{0pt} \tablehead{ \colhead{}
    & \colhead{} & \colhead{} & \colhead{} & \colhead{Mass Scaling} &
    \colhead{$\Mtwo$\tablenotemark{a,b}} \\
    \colhead{Mass Type} & \colhead{Proxy} & \colhead{Measurement} &
    \colhead{Units} & \colhead{Relation} &
    \colhead{($10^{14}$ \msun)}} \startdata 
Dispersion &  {\bf Biweight}  & $1179\, ^{+232}_{-167}$ & km/s &  $\sigma$--$\Mtwo$\citep{evrard08}   & {\boldmath $10.4\,^{+6.1}_{-4.4}$} \\
           & Gapper            & $1170\, ^{+240}_{-128}$ & km/s & $\sigma$--$\Mtwo$ \citep{evrard08}   & $10.1\,^{+6.2}_{-3.3}$ \\
           & Std Deviation     & $1138\, ^{+205}_{-132}$ & km/s & $\sigma$--$\Mtwo$ \citep{evrard08}   & $9.3\,^{+5.0}_{-3.2}$ \\
X-ray      & {\boldmath $Y_X$} &  $ 5.3 \pm 1.0 $ &  $\times 10^{14}$ \msun keV &  $Y_X$--$\Mfive$ \citep{vikhlinin09} & {\boldmath $8.23 \pm 1.21$} \\ 
           & $T_X$ & $ 7.5\, ^{+1.7}_{-1.1}$ & keV & $T_X$--$\Mfive$ \citep{vikhlinin09} & $8.11 \pm 1.89$ \\
SZE        & {\boldmath $Y_{\mbox{\scriptsize \bf SZ}}$} &  $3.5 \pm 0.6$ &  $\times 10^{14}$ \msun keV &  $Y_{\mbox{SZ}}$ -- $\Mfive$ (A10) & {\boldmath $7.19 \pm 1.51$} \\ 
           & S/N at 150 GHz & 7.69 & & $\xi$ -- $\Mfive$ (V10) & $5.03 \pm 1.13 \pm 0.77$ \\
Richness   & {\boldmath $\Ntwobf$} &  $80 \pm 31$ &  galaxies &  $\Ntwo$ -- $\Mtwo$ (H10) & {\boldmath $8.5 \pm 5.7 \pm 2.5$} \\ 
           & $\Ngal$               &  $66 \pm 7$  & galaxies  & $\Ngal$ -- $\Mtwo$ (H10)  & $9.2 \pm 4.9 \pm 2.7 $ \\ \hline \\
Best       & {\bf Combined}       &              &           &                           &   {\boldmath $7.95 \pm 0.92$}
\enddata
\tablenotetext{a}{$\Mfive$ masses were scaled to $\Mtwo$ masses
  assuming an NFW density profile and the mass-concentration relation
  of \citet{duffy08}.}  \tablenotetext{b}{We do not correct the
  dynamical masses for the small potential bias in the dispersion
  discussed in \textsection{\ref{Sec: dynamics}}.}
\end{deluxetable*}

\subsection{Combined Mass for \clname}

Table \ref{Tab: masses} lists total mass measures for \clname\ from
dynamical, X-ray, SZE and optical richness mass proxies.  For each
physical probe, the bold entries indicate the specific proxy expected
to yield the most robust mass.  For dispersions, this is the biweight
measure, which is optimal for the number of spectroscopic members we
have \citep{beers90}.  For the X-ray, the $Y_X$ estimator is chosen
over the $T_X$ estimator because it is expected to be significantly
less biased for mergers and to have less intrinsic scatter with mass
\citep{kravtsov06,vikhlinin09}.  For the SZE, the $\Ysz$ estimator is
chosen over the S/N estimator because scaling relations for it have
been calibrated directly from external observations rather than
simulations and it is expected to have less intrinsic scatter with
mass (\citealt{kravtsov06}; V10).  Finally, the $\Ntwo$ estimator is
the best-tested optical richness measure.

We optimally combine these independent measurements, weighting them by
their errors. We symmetrize the errors on the dynamical mass and add
the statistical and systematic errors in quadrature for the richness.
The combined mass is insensitive to these details since these are the
least constrained measures.  The resulting mass is $\Mtwo = (7.95 \pm
0.92) \times 10^{14}$ \msun, which we take as our best estimate of the
mass of \clname.

\subsection{ \clname\ in a Cosmological Context}

Using the \citet{tinker08} mass function we calculate that \clname,
which at $z=1.067$ is already almost as massive as the Coma cluster,
will increase in mass by a factor of $\sim 4$ over the next 8 Gyr.  It
should therefore grow into one of the most massive clusters in the
Universe by the present day.

Historically, the existence of individual massive galaxy clusters at
high redshift has been used to constrain cosmological models
\citep[\eg][]{donahue98}.  While we are entering an era where large
samples of SZ clusters will soon be available in addressing such
questions, it is interesting to ask whether the existence of \clname\
is expected in our survey volume for a concordance $\Lambda$CDM
cosmology. Convolving the \citet{tinker08} mass function with a
Gaussian mass probability function for the best-estimate mass of
\clname, we find that $\Lambda$CDM predicts the existence of $0.18$
clusters of this mass or higher at $z>1$ in 178 deg$^2$.  We note that
the $\Mfive-Y_X$ relation was calibrated at lower redshifts ($z \la
0.6$) than \clname, and the accuracy of this relation has not been
verified at $z\ga 1$.  Given the caveats and uncertainties inherent in
this calculation, we conclude that the existence of \clname\ in our
survey volume is unsurprising.  The complete SPT sample, combined with
improved mass scaling relations based on complementary mass measures,
will permit $\Lambda$CDM predictions for the high end of the cluster
mass function to be robustly tested at high redshift.

\section{Conclusions}

We report the spectroscopic confirmation \clname\ at $z = 1.067$, the
first SZE-selected galaxy cluster at $z > 1$.  We measure a robust
velocity dispersion from 18 early-type members of $\sigma = 1179\,
^{+232}_{-167}\,$ km/s, corresponding to a dynamical mass of $\Mtwo =
1.0\,^{+0.6}_{-0.4} \times 10^{15}$ \msun.  \clname\ is the most
dynamically massive cluster yet identified, from any method, at $z>1$.

We find excellent consistency across several independent mass
measures.  The measured dispersion and X-ray temperature of \clname\
fall right on the $\sigma$--$T_X$ relation.  The X-ray, SZE and
richness-based mass estimates are all consistent with the dynamical
mass, and with each other, within the errors.  Combining all the mass
measures, we derive a best-estimate mass for \clname\ of $\Mtwo =
(7.95 \pm 0.92) \times 10^{14}$ \msun.  

In $\Lambda$CDM we expect 0.18 clusters consistent with this mass
above $z>1$ in our survey area.  Given the uncertainties in X-ray
scaling relations at high redshift, we conclude that the existence of
\clname\ in our survey volume is unsurprising.  The complete SPT
sample will provide the large sample of clusters required to robustly
test the high redshift, high mass end of the cluster mass function.

\acknowledgments

The South Pole Telescope is supported by the National Science
Foundation through grant ANT-0638937.  Partial support is also
provided by the NSF Physics Frontier Center grant PHY-0114422 to the
Kavli Institute of Cosmological Physics at the University of Chicago,
the Kavli Foundation, and the Gordon and Betty Moore Foundation.  This
work is based in part on observations made with the Spitzer Space
Telescope, which is operated by the Jet Propulsion Laboratory,
California Institute of Technology under a contract with NASA. Support
for this work was provided by NASA through an award issued by
JPL/Caltech.  This paper includes data gathered with the 6.5 meter
Magellan Telescopes located at Las Campanas Observatory, Chile.  This
work is based in part on observations obtained with the Chandra X-ray
Observatory (CXO), under contract SV4-74018, A31 with the Smithsonian
Astrophysical Observatory which operates the CXO for NASA.  We are
very grateful for the efforts of the {\it Spitzer}, {\it Chandra},
Magellan, and CTIO support staff without whom this paper would not be
possible.  Support for M.B. was provided by the W. M. Keck Foundation.
B.S. acknowledges support from the Brinson Foundation.

\end{document}